\documentstyle[12pt,epsf]{article}
\begin{document}
\newcommand{\bq}{\begin{equation}}
\newcommand{\eq}{\end{equation}}
\newcommand{\bqn}{\begin{eqnarray}}
\newcommand{\eqn}{\end{eqnarray}}
\newcommand{\nb}{\nonumber}
\newcommand{\lb}{\label}

\baselineskip 0.75cm

\title{On the sources of static plane symmetric vacuum space-times}
\author{ M.F.A. da Silva \thanks{e-mail address: 
 mfas@symbcomp.uerj.br} \ and 
Anzhong Wang \thanks{e-mail address: wang@symbcomp.uerj.br}\\
\small Departamento de F\' {\i}sica Te\' orica,
Universidade do Estado do Rio de Janeiro, \\
\small Rua S\~ ao Francisco Xavier 524, Maracan\~ a,
20550-013 Rio de Janeiro~--~RJ, Brazil\\
\small and\\
\small Observat\'orio
Nacional~--~CNPq, Rua General Jos\'e Cristino 77,\\
\small S\~ao Crist\'ov\~ao, 20921-400 Rio de Janeiro~--~RJ, Brazil\\
   \\
N. O. Santos \thanks{e-mail address: nos@on.br}\\
\small Departamento de Astrof\'{\i}sica, Observat\'orio
Nacional~--~CNPq, \\  
\small Rua General Jos\'e Cristino 77, S\~ao Crist\'ov\~ao,
20921-400 Rio de Janeiro~--~RJ, Brazil\\
\small and\\
\small Departamento de F\' {\i}sica Te\' orica,
Universidade do Estado do Rio de Janeiro, \\
\small Rua S\~ ao Francisco Xavier 524, Maracan\~ a,
20550-013 Rio de Janeiro~--~RJ, Brazil}

\maketitle

\newpage

\begin{abstract}

\baselineskip 0.9cm

The static vacuum plane spacetimes are considered, which have two
non-trivial solutions: The Taub solution and the Rindler solution.
Imposed reflection symmetry, we find that the source for the Taub
solution does not satisfy any energy conditions, which is consistent
with previous studies, while the source for the Rindler solution
satisfies the weak and strong energy conditions (but not the dominant
one). It is argued that the counterpart of the Einstein theory to the
gravitational field of a massive Newtonian plane should be described by
the Rindler solution, which represents also a uniform gravitational
field.

\end{abstract}

\baselineskip 0.9cm


\newpage

It is well-known that the unique solution of vacuum static plane
symmetric space-time with {\em non-null curvature} is the Taub plane
solution \cite{Taub1}, which in the usual plane symmetric coordinates,
$x^\mu=\{T, Z, X, Y\}$ ($\mu=0, 1, 2, 3$), is given by
\begin{equation}
\label{1}
ds^2={_1\over Z^{2/3}}dT^2-dZ^2-Z^{4/3}(dX^2  +dY^2).
\end{equation}
As shown by Taub himself, the space-time is asymptotically flat as
$|Z|\rightarrow +\infty$ and singular at $Z=0$. It is generally
believed that the singularity should be replaced by a regular source for
more realistic models. It is this belief that attracted much of
attention in finding sources for the Taub solution \cite{Bonnor}. In
particular, Dolgov and Khriplovich \cite{DolgovKhriplovich} showed that
static space-times with reflection plane symmetry are not free of
space-time singularity, provided that the energy density of the sources
is positive. This implies that the sources of Taub's solution  have
negative energy density, if we require that the space-time be static,
reflection symmetric, and free of space-time singularity.

On the other hand, from the work of Vilenkin \cite{Vilenkin} and Novotn\'y
et al \cite{Novotnyetal} we can see that the Newtonian limit of such
space-times exists. As a matter of fact, it takes the form
\begin{equation}
\label{2}
ds^2=(1+8\pi\rho |Z|) dT^2-(1-8\pi\rho |Z|) (dZ^2+dX^2+dY^2),
\end{equation}
which satisfies the linearized Einstein's field equations
$\Box\Psi_{\mu\nu}=8\pi\rho \delta(Z) {\delta_\mu}^T{\delta_\nu}^T$,
where $\delta(Z)$ denotes the Dirac delta function, and $\rho$ the
surface energy density of the plane $Z = 0$.  The results obtained in
\cite{Bonnor,DolgovKhriplovich} and the ones obtained in
\cite{Vilenkin,Novotnyetal} seemingly contradict: There exists Newtonian
limit but does not exist exact solutions to the Einstein field
equations.

In this Letter, we shall show that there
is no contradiction. In fact, exact solutions with positive mass indeed exist
but the space-time outside sources is not Taub's, rather than that
of Rindler \cite{Rindler}, which represents an uniform
gravitational field. Recall that the gravitational field of a massive
plane in Newtonian theory is also uniform!

To show the above, let us start with the general form of the metric for
the static plane symmetric space-times \cite{Taub1}
\begin{equation}
\label{3}
ds^2=f(z)(dt^2-dz^2)-g(z)(dx^2+dy^2),
\end{equation}
where $\{x^\mu\}\equiv\{t, z, x, y\}$ is another set of plane symmetric
coordinates.

The non-vanishing components of the Einstein tensor for the above metric
are given by
\bqn
\label{4}
G_{00} &=& \frac{1}{4fg^2}\left(f{g'}^2 - 4fgg'' + 2gf'g'\right),\nb\\
G_{11}&=& \frac{1}{4fg^2}\left(2gf' +fg'\right)g',\nb\\
G_{22}&=& G_{33}=\frac{1}{4gf^3}\left(2fg^2f''-2g^2{f'}^2 + 2gf^2g''
-f^2{g'}^2\right),
\eqn
where a prime denotes the ordinary differentiation with respect to $z$.

>From the expression of $G_{11}$, we can see that the vacuum solutions of
the Einstein's field equations can be divided into three different
cases:
$$
i)\; 2gf' +fg' =0, \;\; g' \neq 0;\;\; 
ii)\; 2gf' + fg'\neq 0,\;\; g' =0;
iii)\; f' = 0 = g'.
$$
In Case i), using equations (\ref{4}) it is easy to show that
the general Einstein vacuum solution is given by
\begin{equation}
\label{7}
f(z)={\alpha^2\over\sqrt{z+\gamma}}, \quad g(z)=\beta^2(z+\gamma),
\end{equation}
where $\alpha$, $\beta$ and $\gamma$ are integration constants. Setting
$$
T=\left({4\alpha^4\over 3}\right)^{1/3} t,
\quad Z={4\alpha\over 3}(z+\gamma)^{3/4},\\
X=\beta\left({3\over 4\alpha}\right)^{2/3} x,\quad 
Y=\beta\left({3\over 4\alpha}\right)^{2/3} y,
$$
the corresponding solution will take the same form as that given by
equation (\ref{1}). Therefore, Case i) gives exactly the Taub solution.

In Case ii), it can be shown that the Einstein vacuum equations have the
general solution
\begin{equation}
\label{8}
f(z)=e^{az+b},\quad g(z)=c^2,
\end{equation}
where $a$, $b$ and $c$ are arbitrary constants. Introducing the new
coordinates
$$
T={a\over 2}t, \quad Z={2\over a}e^{(az+b)/2},\quad  X=cx, \quad Y=cy,
$$
the corresponding metric will take the form
$$
ds^2=Z^2dT^2-dZ^2-(dX^2+dY^2),
$$
but this is the Rindler solution \cite{Rindler}, which represents an
uniform gravitational field, and the corresponding 
Riemann tensor is identically zero.

In Case iii), the vacuum Einstein field equations have the trivial
solution, $f= Const. $ and $g=Const.$ Clearly, this
exactly corresponds to the Minkowski space-time.

Therefore, in general, the static Einstein vacuum field equations with
plane symmetry have two non-trivial classes of solutions. To find
sources for these two classes of solutions, we shall model the sources
as infinitely thin planes.  The main reason for doing so is because in
this case the Newtonian theory has an unique solution. Moreover, as
showed in \cite{Vilenkin,Novotnyetal}, its Newtonian limit also
exists. To construct such sources, we shall use the ansatz,
\begin{equation}
\label{9}
z\rightarrow|z|
\end{equation}
in the solutions (\ref{7}) and (\ref{8}), respectively. Such a
substitution  is physically equivalent first to cut
a space-time into two regions, $z>0 $ and $z<0$, and then join the
region $z>0$ with a copy of it, so that the resulted space-time has a
reflection symmetry with respect to the surface $z=0$. Clearly, this is
possible only when a matter shell appears on the surface, which serves
as a mirror of the gravitational field.

The form of the matter shell can be calculated using Taub's formulae
\cite{Taub2}.  Following Taub \footnote{The formulae given by Taub are
equivalent to those of Israel \cite{Israel}, when the matching
hypersurface is spacelike or timelike, as shown by Taub himself
\cite{Taub2} (See also \cite{wangect}).}, we first introduce a 
surface tensor $b_{\mu\nu}$ via the
relations
\begin{equation}
\label{10} 
[g_{\mu\nu,\lambda}]^- ={g^{+}_{\mu\nu,\lambda}}|_{z=0^{+}} - {g^{-}_{\mu\nu,\lambda}}|_{z=0^{-}} = \eta_\lambda b_{\mu\nu},
\end{equation}
where $\eta_\lambda$ is the normal vector to the hypersurface $z=0$, and
is given by $\eta_\lambda={\delta^z}_\lambda$. $g^{+}_{\mu\nu} \;
(g^{-}_{\mu\nu})$ are the metric coefficients given in the region $z > 0
\; (z < 0)$.  Once $b_{\mu\nu}$ is known, the surface energy-momentum
tensor $\tau_{\mu\nu}$ can be read off from the expression
\begin{equation}
\label{11}
\tau_{\mu\nu}={1\over 16\pi}\{b(\eta g_{\mu\nu}-
\eta_\mu\eta_\nu)+\eta_\lambda(\eta_\mu 
{b_\nu}^\lambda+\eta_\nu{b_\mu}^\lambda)-
(\eta b_{\mu\nu}+\eta_\lambda\eta_\delta 
b^{\lambda\delta}g_{\mu\nu})\},
\end{equation}
where $\eta\equiv \eta_\lambda\eta^\lambda$ and $ b\equiv
{b_\lambda}^\lambda$. In terms of $\tau_{\mu\nu}$ the energy-momentum
tensor $T_{\mu\nu}$ of the four-dimensional space-time takes the form
$T_{\mu\nu}={T^{+}_{\mu\nu}}H(z)+{T^{-}_{\mu\nu}}H(z) +
\tau_{\mu\nu}\delta(z)$, where $H(z)$ denotes the Heaviside function,
which is  one for $z \ge 0$ and zero for $z<0$.  ${T^{\pm}_{\mu\nu}}$
are the energy-momentum tensor calculated respectively in the regions
$z>0$ and $z<0$. In the present case, since the space-time is vacuum in
these two regions, we have ${T^{\pm}_{\mu\nu}} =0$.

For the general metric (\ref{3}), it can be shown that $\tau_{\mu\nu}$
can be written as
\begin{equation}
\label{12}
\tau_{\mu\nu}=\rho u_\mu u_\nu +p(x_\mu x_\nu+y_\mu y_\nu),
\end{equation}
where

\begin{equation}
\label{13}
\rho=-{[g^\prime]^-\over 8\pi fg},\quad
 p={1\over 16\pi f^2 g}[(fg)^\prime]^-,
\end{equation}
and
\begin{equation}
\label{14}
u_\mu\equiv \sqrt{f}{\delta_\mu}^t,\quad 
x_\mu\equiv \sqrt{g}{\delta_\mu}^x,\quad  
y_\mu\equiv \sqrt{g}{\delta_\mu}^y,
\end{equation}
with $[G]^-\equiv G^+|_{z=0^+}-G^-|_{z=0^-}$.  Clearly, $\rho$
represents the surface energy density of the shell, and $p$ the pressure
in the $dx$ and $dy$ directions, measured by observers who are at rest
with respect to the shell.

Substituting equation (\ref{9}) into equation (\ref{7}), we find
\begin{equation}
\label{15}
f(z)={\alpha^2\over\sqrt{|z|+\gamma}},\quad g(z)=\beta^2(|z|+\gamma).
\end{equation}
As mentioned previously, Taub's solution is singular on $z+\gamma=0$. To
avoid this singularity in the solution (\ref{15}), we require
$\gamma>0$. Clearly, the spacetime in the region $z>0$ is locally
isometric to Taub's with $Z>Z_0$, where $Z_0\equiv
(4\alpha/3)\gamma^{3/4}$, while in the region $z<0$, it is a copy of the
spacetime of the region $z>0$. On the hypersurface $z=0$, the surface
energy-momentum tensor is given by equation (\ref{12}) with $\rho$ and
$p$ now being given by
\begin{equation}
\label{16}
\rho=-4p=-{1\over 4\pi\alpha^2\sqrt{\gamma}}.
\end{equation}
Clearly, now the surface energy density of the shell is negative and
does not satisfy any of the three energy conditions, weak, dominant and
strong \cite{HawkingEllis}. This is consistent with the results obtained
in \cite{Bonnor, DolgovKhriplovich}. It is interesting to note that
$\rho$ and $p$ both depend on $\alpha^2\sqrt{\gamma}$, 
which characterizes the strength of the
energy density per unit area of the shell. This is understandable, if we
recall that its Newtonian counterpart also has one independent parameter,
although sometimes people tend to believe that Taub's solution
essentially has no physically independent free parameters.

On the other hand, substituting equation (\ref{9}) into equation 
(\ref{8}) we find
\begin{equation}
\label{17}
f(z)=e^{a|z|+b}, \quad g(z)=c^2.
\end{equation}
Then, inserting these expressions into equation (\ref{13}) we obtain
\begin{equation}
\label{18}
\rho=0, \quad p={ae^{-b}\over 8\pi}.
\end{equation}
Thus, in this case the energy density of the shell measured by those
observers who are comoving with the shell is zero, while the pressure in
the $dx$ and $dy$ directions in general is not. This is a little bit
strange. The reason  for it is that the shell does not satisfy the
dominant energy condition, as one can easily show from equation
(\ref{18}). However, it does satisfy the weak and strong energy
conditions, provided $a>0$. Moreover, only the comoving observers
measure zero surface energy density. Other observers usually do not. For
example, observers with the four-velocity
$v_\mu=\sqrt{(1+\delta^{2})f}{\delta_\mu}^t+ \delta
\sqrt{g}{\delta_\mu}^x$ will measured the density as
$\rho_v=T_{\mu\nu}v^\mu v^\nu= \delta^{2}p\geq 0$, where $\delta$ is an
arbitrary constant. It should be noted that our results do not
contradict with the ones obtained in \cite{DolgovKhriplovich}. In fact,
the function $u$ defined there is a constant in the present case, a case
that is excluded in \cite{DolgovKhriplovich}.

As showed above, the most general non-trivial static plane vacuum
solutions are the Taub solution and the Rindler solution, and the
sources of the Taub solution always have negative energy density.
Therefore, the unique solution that corresponds to the Newtonian limit
(\ref{2}) is the solution given by equation (\ref{17}). The
space-time outside the source represents an uniform gravitational field,
and is locally isometric to the Rindler solution \cite{Rindler}. The
coordinate transformations that bring the solution of (\ref{17}) (for $p$
being very small) to the one of equation (\ref{2}) are complicate,
and we have not found them, yet.

It is well-known that Rindler space-time has a non-trivial topology. In
particular, it has event horizons. Thus, it would be very interesting to
study the global structure of the solution (\ref{17}). To do this, let us
first consider the following coordinate transformations
\bqn
\label{25}
\bar T &=& {2\over a}e^{az+b/2}{\rm sinh}{a t\over 2},\;\;
\bar Z = {2\over a}e^{az/2}{\rm cosh}{a t\over 2},\nb\\
\bar X &=& cx,\;\;
\bar Y = cy,\; (z \ge 0).
\eqn
Then we find that the metric takes exactly the Minkowski form $ds^2=d\bar
T^2-d\bar Z^2-d\bar X^2-d\bar Y^2$ in this region. From equation (\ref{25})
 we obtain
\begin{equation}
\label{26}
\bar Z^2-\bar T^2=4/a^2, \quad (z \ge 0).
\end{equation}
Equations (\ref{25}) and (\ref{26}) show the mapping between the
($t$,$z$)-plane and the ($\bar T$,$\bar Z$)-plane. In particular, the
half plane $z>0$ is mapped to region $I$ where $\bar
Z>(4a^{-2}e^{az+b}+\bar T^{2})^{1/2}$ [cf. Fig. 1], while the
hypersurface $z=0$ is mapped to one branch of the hyperbole $\bar
Z^2-\bar T^2={4a^{-2}}e^b (\bar Z>0)$. Region $I^\prime$ [where $\bar
Z<-\left({4a^{-2}}e^{az+b}+\bar T^2)^{1/2}\right]$ is an extended
region, while the region $-\left({4a^{-2}}e^{az+b}+\bar T^2\right)^{1/
2}< \bar Z<\left({4a^{-2}}e^{az+b}+\bar T^2\right)^{1/ 2}$ corresponds
to $z<0$ and does not belong to the space-time described by equation
(\ref{17}). Note that the hypersurfaces $\bar T^2=\bar Z^2$ represent
the event horizons of the Rindler space-time \cite{Rindler}, which is
excluded from the space-time.

On the other hand, in region $z\leq 0$, if we make the same coordinate
transformations (\ref{25}) but with $z$ being replaced by $-z$, the
corresponding solution (\ref{17}) also takes the Minkowski form but now
with
\begin{equation}
\label{27}
\tilde Z^2-\tilde T^2={4\over a^2}e^b,\quad (z \le 0). 
\end{equation}
Thus, the global structure of this region is similar to that with $z\ge
0$.  In particular, the half plane $z<0$ is mapped to region $II$
where $\tilde Z>\left({4a^{-2}}e^{az+b}+\tilde T^2\right)^{1/2}$ [cf.
Fig. 2], while the hypersurface $z=0$ is mapped to the branch of the
hyperbole $\tilde Z^2-\tilde T^2={4a^{-2}}e^b,\quad \tilde Z>0$. Region
$II'$ where $\tilde Z\leq-\left({4a^{-2}}e^{-az+b}+\tilde
T^2\right)^{1/2}$ is an extended region in the Minkowski coordinates
$\{\tilde T, \tilde Z, \tilde X, \tilde Y\}$. The global structure of
the whole space-time is illustrated by Fig. 2. From there we can see
that there essentially exist two plane shells  $\overline{ab}$ and
$\overline{cd}$. Each of the two shells connected two flat regions, in
which the gravitational field is uniform, and these two regions are
geodesically complete and free of any event horizons.

\section*{Acknowledgment} 

The authors gratefully acknowledge financial assistance from CNPq.

\newpage

\section*{Figure Captions}

{\bf Fig. 1} The spacetime project for the region $z \ge 0$ in the ($\bar{T},
\bar{Z}$)-plane for the solution (\ref{17}). Region $I$ corresponds to
$z > 0$, while region $I'$ is an extended region in the Minkowski-like
coordinates, $\bar{T}, \bar{Z}, \bar{X}$ and $\bar{Y}$.

\vspace{.6cm}

\noindent {\bf Fig. 2} The corresponding Penrose diagram for the solution
(\ref{17}). $\overline{ab}$ and $\overline{cd}$ represent two thin matter
shells, each of which connects two asymptotically flat regions. Therefore,
each of the two diamonds represents a geodesically complete spacetime.

\end{document}